\documentclass[a4paper]{article}

\usepackage{INTERSPEECH2020}
\usepackage{subfig}
\usepackage{bm}
\usepackage{float}

\def\ATT{\mathrm{ATT}}
\def\FF{\mathrm{FF}}
\def\Q{\mathrm{Q}}
\def\K{\mathrm{K}}
\def\V{\mathrm{V}}
\def\H{\mathrm{H}}

\title{When Can Self-Attention Be Replaced by Feed Forward Layers?}
\name{Shucong Zhang, Erfan Loweimi, Peter Bell, Steve Renals}
\address{
 Centre for Speech Technology Research, University of Edinburgh, Edinburgh, UK}
\email{s1603602@sms.ed.ac.uk, \{e.loweimi, 
 peter.bell, s.renals\}@ed.ac.uk}

\begin{document}

\maketitle
\begin{abstract}
Recently, self-attention models such as Transformers have given competitive results compared to recurrent neural network systems in speech recognition. The key factor for the outstanding performance of self-attention models is their ability to capture temporal relationships without being limited by the distance between two related events. However, we note that the range of the learned context progressively increases from the lower to upper self-attention layers, whilst acoustic events often happen within short time spans in a left-to-right order. This leads to a question: for speech recognition, is a global view of the entire sequence still important for the upper self-attention layers in the encoder of Transformers?  To investigate this, we replace these self-attention layers with feed forward layers. In our speech recognition experiments (Wall Street Journal and Switchboard), we indeed observe an interesting result: replacing the upper self-attention layers in the encoder with feed forward layers leads to no performance drop, and even minor gains. Our experiments offer insights to how self-attention layers process the speech signal, leading to the conclusion that the lower self-attention layers of the encoder encode a sufficiently wide range of inputs, hence learning further contextual information in the upper layers is unnecessary.



\end{abstract}

\section{Introduction}

Self-attention networks (SANs) has recently become a popular research topic in the speech recognition community \cite{Dong2018Speech-transformer, zhou2018syllable, pham2019veryDeep, povey2018time, zeyer2019comparison}. Previous studies showed SANs can yield superior speech recognition results compared to recurrent neural networks (RNNs) \cite{wang2019transformerHybrid, karita2019comparative}. 

RNNs are conventional models to model sequential data. However, due to gradient vanishing, it is difficult for RNNs to model long-range dependencies \cite{bengio1994learning}. Although gated structures, such as Long Short-Term Memory (LSTM)~\cite{hochreiter1997LSTM} and Gated Recurrent Unit (GRU)~\cite{chung2014GRU} are proposed to alleviate this problem, capturing temporal relationships across a wide time span remains challenging for these models. In SANs, self-attention layers encode contextual information through attention mechanisms \cite{bahdanau2015neural,luong2015effective,vaswani2017attention}. With the attention mechanism, when learning the hidden representation for each time step of a sequence, a self-attention layer has a global view of the entire sequence and  it can thus capture temporal relationships without the limitation of range. This is believed to be a key factor for the success of SANs \cite{vaswani2017attention}. 

In this paper we study Transformers~\cite{vaswani2017attention}, end-to-end SAN-based models with two components: an encoder and a decoder. The encoder uses self-attention layers to encode input sequences. At decoding time step $t$, the decoder generates the current output by attending to the encoded input sequence and to the outputs generated before time $t$. For attention-based RNN end-to-end models \cite{luong2015effective, bahdanau2015neural}, the RNN encoder encodes the input sequence. The RNN decoder interacts with the encoded input sequence through an attention layer to produce outputs. 

Previous work on attention-based RNN end-to-end models has shown that for speech recognition, since acoustic events usually happen in a left-to-right order within small time spans, restricting the attention to be monotonic along the time axis improves the model's performance \cite{tjandra2017local, kim2017joint, zhang2019windowed}. This seemingly leads to a  contradiction to the reason of SANs' success: if the global view provided by the attention module of self-attention layers is beneficial, then why does forcing the attention mechanism to focus on local information result in performance gains for RNN end-to-end models?

To investigate this, we explore replacing the upper (further from the input) 
self-attention layers of the Transformer's encoder with feed forward layers. We ran extensive experiments on a read speech corpus Wall Street Journal  (WSJ) \cite{paul1992wsj} and a conversational telephone  speech corpus Switchboard (SWBD) \cite{godfrey1992switchboard}. We found that replacing the upper self-attention layers with feed forward layers does not yield higher error rates -- it even leads to improved accuracies. Since a feed forward layer can be viewed as a pure ``monotonic left-to-right diagonal attention'', this observation does not contradict the previous studies on RNN-based end-to-end models which restrict the attention to be diagonal. Thus, it indicates the inputs to the upper layers of the Transformer encoder have encoded enough contextual information and learning further temporal relationships through self-attention is not helpful. These experiments also do not nullify the benefits of the self-attention layers -- the range of learned context is increased from bottom to up through the self-attention layers and it is important for the lower layers to well encode the context information. Only when the lower layers have captured sufficient contextual information, the attention mechanism becomes redundant for the upper layers. 

It should be noticed that for the attention-based RNN models, the attention mechanism interacts with both the decoder and the encoder. Since an output unit (e.g. a character) is often related to a small time span of acoustic features, the attention needs to attend a small window of the elements in the encoded input sequence in a left-to-right order. In this work we study the self-attention encoder which learns the hidden representation for each time step of the input sequence. Thus, feed forward layers, which can be viewed as ``left-to-right attention without looking at the context'', are sufficient in learning further abstract representation for the frame in the current time step when temporal relationships among input frames are well captured.

Our observations also make practical contributions. Compared to self-attention layers,  feed forward layers have a reduced number of parameters. Furthermore, without parallel computation, the time complexity for a self-attention layer is $o(n^2)$ where $n$ is the length of the input sequence. Replacing self-attention layers with feed forward layers also reduces the time complexity. 

\section{Multi-head attention}
Self-attention and its multi-head attention module \cite{vaswani2017attention} which uses multiple attention mechanisms to encode context are key components of Transformers. Previous works have analysed the importance of these modules. For a self-attention layer, a single-layer feed forward module is stacked on the multi-head attention module. Irie et al~\cite{irie2020how} extend the single-layer feed forward module to a multi-layer module, arguing it can bring more representation power, and show that a SAN with fewer modified self-attention layers can have minor performance drops compared to a SAN with a larger number of the original self-attention layers. However, with fewer layers, the models with the modified self-attention layers give reduced number of parameters/decoding time. In this work we study the effect of the stacked context among the self-attention layers of the encoder.  We do not change the architecture of the self-attention layers and we replace the upper self-attention layers in the encoder of Transformers with feed forward layers.

Michel et al~\cite{michel2019are} remove a proportion of the heads in the multi-head attention for each self-attention layer in Transformers, finding it leads to minor performance drops. This implies the benefits of multi-head attention mainly come from the training dynamics. In our work, instead of removing some attention heads, we replace the entire self-attention layer with feed forward layers to investigate how the self-attention layers encode the speech signal.


When the upper self-attention layers are replaced with feed forward layers, the architecture of the encoder is similar to the CLDNN -- Convolutional, Long Short-Term Memory Deep Neural Network \cite{sainath2015CLDNN}. The CLDNN uses an LSTM to model the sequential information and a deep neural network (DNN) to learn further abstract representation for each time step. Stacking a DNN on an LSTM results in a notable error rate reduction compared to pure LSTM models. While we found the upper self-attention layers of the encoder of Transformers can be replaced with feed forward layers, stacking more feed forward layers does not result in further performance gains. The main goal of this work is to understand the self-attention encoder.

\section{Model Architecture}
In this section we describe multi-head attention, self-attention layers, and the self-attention encoder module in Transformers. Then, we introduce the replacement of the upper self-attention layers of the encoder by feed forward layers. 

\subsection{Multi-head Attention}
Multi-head attention uses attention mechanisms to encode sequences \cite{vaswani2017attention}. We firstly consider a single attention head. The input sequences to the attention mechanism are mapped to a query sequence $\bm{Q}$, a key sequence $\bm{K}$ and a value sequence $\bm{V}$. These sequences have the same dimension, and $\bm{K}$ and $\bm{V}$ have the same length. For the $i$-th element $\bm{Q}[i]$ of $\bm{Q}$, an attention vector is generated by computing the similarity between $\bm{Q}[i]$ and each element of $\bm{K}$. Using the attention vector as weights, the output is a weighted sum over the value sequence $\bm{V}$. Thus, an attention head $\mathrm{A}$ of the multi-head attention can be described as:
\begin{equation}
    \mathrm{A}(\bm{X}^\Q, \bm{X}^\K, \bm{X}^\V) = \mathrm{softmax}(\frac{\bm{Q} \bm{K}^T }{\sqrt{d^{\ATT}}})\bm{V}
\end{equation}
and
\begin{equation}
(\bm{Q} ,\bm{K} ,\bm{V}) = (\bm{X}^\Q \bm{W}^\Q, \bm{X}^{\K}\bm{W}^{\K},\bm{X}^\V \bm{W}^\V),
\end{equation}
where $\bm{X}^\Q \in \mathbb{R}^{ n \times d^{\ATT}}$, $\bm{X}^\K, \bm{X}^\V \in \mathbb{R}^{ m \times d^{\ATT}}$ are inputs and $m,n$ denote the lengths of the input sequences; $\bm{W}^\Q, \bm{W}^\K, \bm{W}^\V \in \mathbb{R}^{ d^{\ATT} \times d^{\ATT}}$ are trainable matrices. The three input sequences $(\bm{X}^\Q, \bm{X}^\K, \bm{X}^{\V})$ can be the same sequence, e.g., the speech signal to be recognised. The multi-head attention $\mathrm{MHA}$ uses $h$ attention heads $(A_{0},A_{1},\cdots,A_{h-1})$ and a trainable matrix $\bm{U}^\H \in \mathbb{R}^{d^{\H} \times d^{\ATT}}, d^{\H} =h \times d^{\ATT}$ to combined the outputs of each attention head:
\begin{equation}
    \mathrm{MHA}(\bm{X}^\Q, \bm{X}^\K, \bm{X}^\V) = (A_{0},A_{1},\cdots,A_{h-1}) \ \bm{U}^\H
\end{equation}

\begin{figure}[tb]
\centering
\subfloat[SA]{\includegraphics[width=0.50\linewidth]{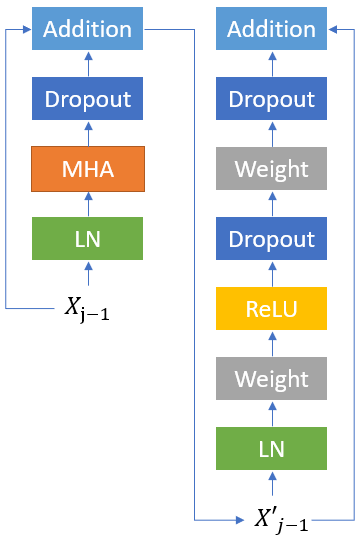}} \quad
\subfloat[FF]{\includegraphics[width=0.222\linewidth]{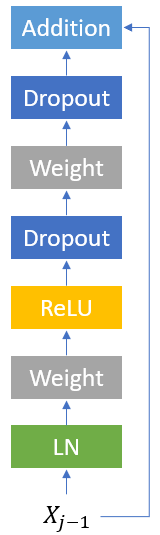}}
\caption{Architectures of a self-attention (SA) encoder layer and a feed forward (FF) encoder layer. LN is layer normalization \cite{ba2016layer}. We omit LN and dropout \cite{srivastava2014dropout} in the equations of encoder layers but they are applied in the experiments.  } 
\label{fig:FF}
\end{figure}
\vspace{-3mm}

\subsection{Self-attention Encoder}
The self-attention encoder in a Transformer is a stack of self-attention layers. The $j$-th 
layer reads the output sequence $\bm{X}_{j-1}$ from its lower layer and uses multi-head attention to process the input sequence. That is,  $(\bm{X}^Q, \bm{X}^K, \bm{X}^V)= (\bm{X}_{j-1},\bm{X}_{j-1},\bm{X}_{j-1})$. The multi-head attention only contains linear operations. Thus, in a self-attention layer, a non-linear feed forward layer is stacked on the multi-head attention module. A self-attention layer in the encoder of a Transformer can be described as:
\begin{equation}
    \bm{X}'_{j-1}= \bm{X}_{j-1} + \mathrm{MHA}(\bm{X}_{j-1},\bm{X}_{j-1},\bm{X}_{j-1})
\end{equation}
\begin{equation}
    \bm{X}_{j} = \bm{X}'_{j-1} + \mathrm{ReLU}(\bm{X}'_{j-1}\bm{S} +\bm{b})\bm{V}+\bm{r}
\end{equation}
where $\bm{S} \in \mathbb{R}^{d^{\ATT}\times d^{\FF}}$, $\bm{V} \in \mathbb{R}^{d^{\FF} \times d^{\ATT}}$, $\bm{b} \in \mathbb{R}^{d^{\FF}}$ and $\bm{r} \in \mathbb{R}^{d^{\ATT}}$. $\bm{S}, \bm{V}, \bm{b}, \bm{r}$ are trainable matrices and vectors . 

\begin{figure*}[h]
\centering
\subfloat[Averaged attention vectors of each attention head of encoder self-attention layer 12 ]{\includegraphics[width=1.33\columnwidth]{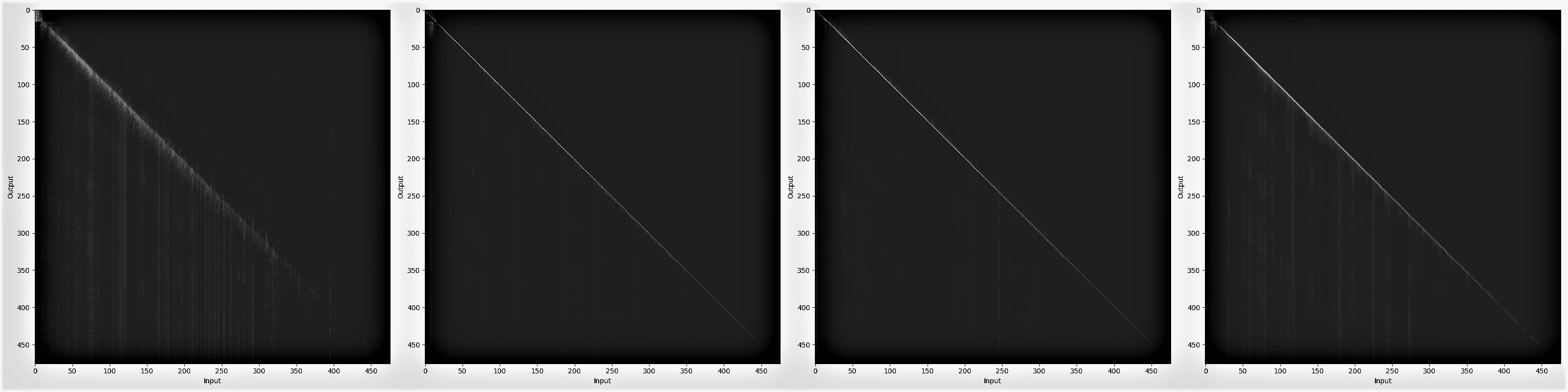}}
\quad
\subfloat[Averaged attention vectors of each attention head of encoder self-attention layer 5 ]{\includegraphics[width=1.33\columnwidth]{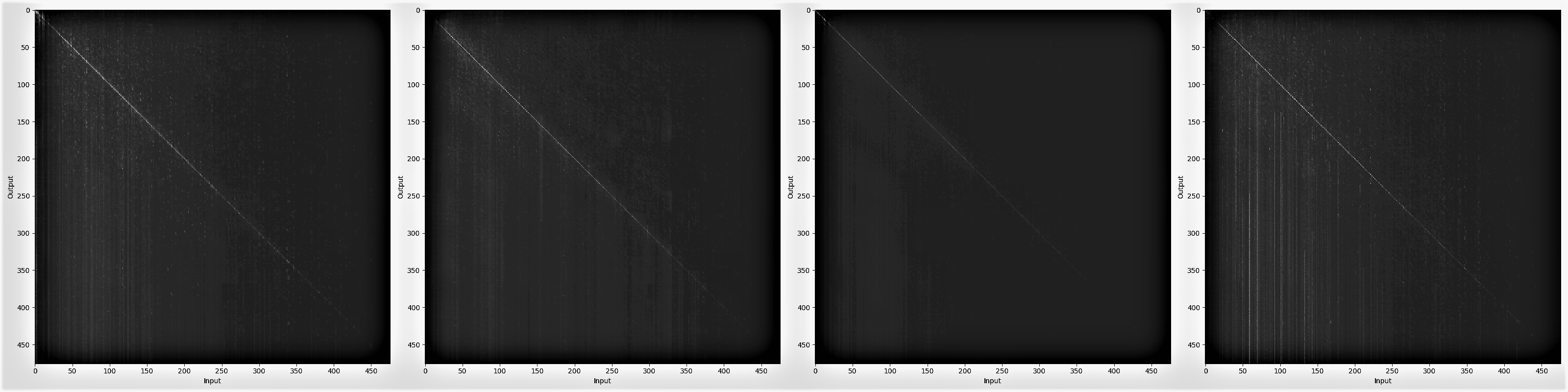}}
\quad
\subfloat[Averaged attention vectors of each attention head of encoder self-attention layer 1]{\includegraphics[width=1.33\columnwidth]{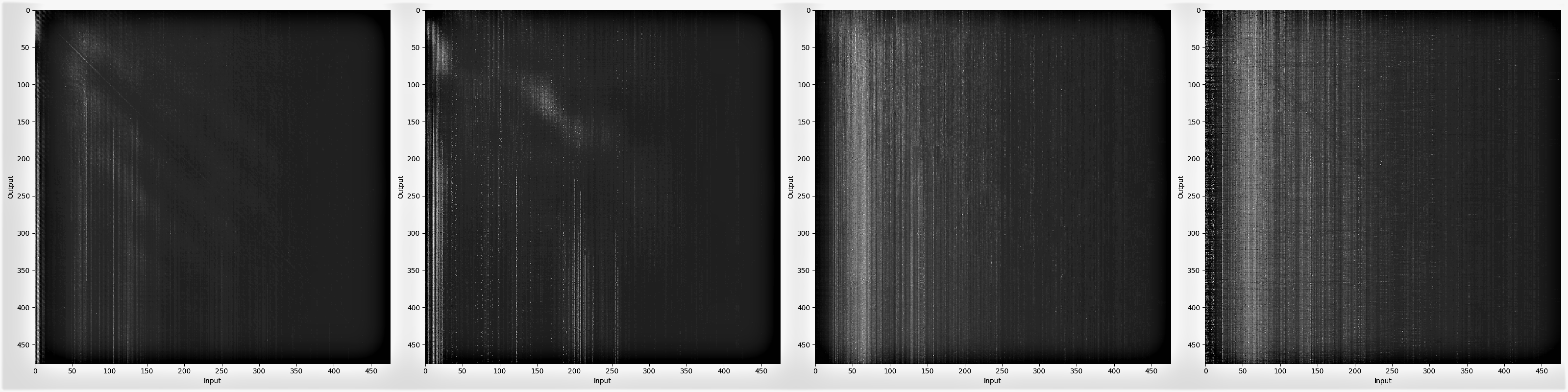}}

\caption{The averaged attention vectors of encoder self-attention layers generated by the baseline Transformer with a 12-layer encoder. The average is over the utterances of WSJ dev93. While the lowest layer (layer1, near input) attends a wide range of context, the middle layer focus more on the local information and the upmost layer almost assign all the attention weights along the diagonal.} 
\label{fig:attention}
\end{figure*}

\subsection{Feed Forward Upper Encoder Layers}
We argue that for the encoder, the upper self-attention layers can be replaced with feed forward layers. In the encoder, since each self-attention layer learns contextual information from its lower layer, the span of the learned context increases from the lower layers to the upper layers. Since acoustic events often happen within small time spans in a left-to-right order, if the inputs to the upper layer have encoded a sufficient large span of context, then it is unnecessary for the upper layers to learn further temporal relationships. Thus, the multi-head attention module which extracts the contextual information may be redundant, and the self-attention layer is not essential.  
However, if  the upper layers of the encoder are self-attention layers, and the lower layers have already seen a sufficiently wide context, then since no further contextual information is needed the attention mechanism will focus on a narrow range of inputs. Assuming that acoustic events often happen left-to-right, the attention vector will tend to be diagonal. Then, $\mathrm{MHA}(\bm{X}_{j-1},\bm{X}_{j-1},\bm{X}_{j-1})\approx \bm{X}_{j-1}$ and self-attention is not helpful, and replacing them with feed forward layers will not lead to a performance drop. The architecture of the feed forward layers is:
\begin{equation}
    \bm{X}_{j} = \bm{X}_{j-1} + \mathrm{ReLU}(\bm{X}_{j-1}\bm{S} +\bm{b}) \ \bm{V}+\bm{r}
\end{equation}
Figure~\ref{fig:FF} demonstrates the architecture of a self-attention layer and a feed forward layer. Furthermore, the feed forward layer can be viewed as the a self-attention layer with a diagonal attention weight matrix -- in the attention weight matrix, the elements among the diagonal are ones and all other elements are zeros. 

\vspace{-2mm}

\section{Experiments and Discussion}

\subsection{Experimental Setup}

We experiment on two datasets, Wall Street Journal (WSJ) which contains 81 hours of read speech training data and Switchboard (SWBD), which contains 260  hours of  conversational telephone speech training data. We use WSJ dev93 and eval92 test sets, and SWBD are eval2000 SWBD/callhome test sets. We use Kaldi \cite{povey2011kaldi} for data preparation and feature extraction -- 83-dim log-mel filterbank frames with pitch \cite{ghahremani2014pitch}. The output units for the WSJ experiments are 26 characters, and the  apostrophe, period, dash, space, noise and \emph{sos} / \emph{eos} tokens. The output tokens for SWBD experiments are tokenized using Byte Pair Encoding (BPE)~\cite{sennrich2016BPE}.

We compare Transformers with different types of encoders. The baseline Transformer encoders comprise self-attention layers and are compared with Transformers whose encoders have feed forward layers following the self-attention layers. Each self-attention/feed forward layer is counted as one layer and encoders with the same number of layers are compared. Except the number of self-attention/feed forward layers in the encoder, all the components of all the models have the same architecture. In each model, below the Transformer's encoder there are two convolutional neural network layers with 256 channels, stride size 2 and kernel size 3, which maps the dimension of the input sequence to $d^{\ATT}$. The multi-head attention components of the self-attention layers have 4 attention heads and $d^{\ATT}=256$. For the feed forward module of the self-attention layers as well as the proposed feed forward encoder layers, $d^{\FF}=2048$. Dropout rate $0.1$ is used when dropout is applied. The decoder of the Transformer has 6 layers. The input sequences to the encoder and the decoder are concatenated with sinusoidal positional encoding \cite{vaswani2017attention}. All models are implemented through the ESPnet toolkit \cite{watanabe2018espnet} and PyTorch \cite{paszke2017automatic}.

Adam \cite{kingma2015adam} is used as the optimizer. The training schedule (warm up steps, learning rate decay) follows previous work \cite{Dong2018Speech-transformer}. The batch size is 32. Label smoothing with smoothing weight 0.1 is used. We train the model for 100 epochs and the averaged parameters of the last 10 epochs are used as the parameters of the final model \cite{Dong2018Speech-transformer}. Besides the loss from the Transformer's decoder $L^{\mathrm{D}}$, a connectionist temporal classification (CTC) \cite{graves2006connectionist} loss $L^{\mathrm{CTC}}$ is also applied to the Transformer encoder \cite{kim2017joint}. With $\lambda = 0.3$ for WSJ and $\lambda = 0.2$ for SWBD, the final loss $L$ for the model is: 
\vspace{-2mm}

\begin{equation}
    L = (1-\lambda)L^{\mathrm{D}} + \lambda L^{\mathrm{CTC}}
\end{equation}
\subsection{Experiments on WSJ}

For the experiments on WSJ, we first train a baseline model with a 12-layer self-attention encoder. Then, we use this model to decode WSJ dev93 and have computed the averaged attention vectors on dev93 generated by the lowest layer (near input), a middle layer and the topmost layer. Figure~\ref{fig:attention} shows the plots of the averaged attention vectors for each attention head of these layers. The lowest layer attends to a wide range of context. The middle layers put more attention weights among the diagonal and the middle two heads of the topmost layer have nearly pure diagonal attention matrices. This implies since the range of learned context 
increases higher up the encoder,
the global view becomes less important and the upper layers focus more on local information. Eventually no additional contextual information is needed so the attention becomes diagonal. For the topmost layer, $\mathrm{MHA}(\bm{X}_{j-1},\bm{X}_{j-1},\bm{X}_{j-1})\approx \bm{X}_{j-1}$.

\vspace{-2mm}
\begin{table}[h]
\caption{Character error rate (CER) on WSJ for the Transformer models with different encoders.}
\vspace{-2mm}
\label{tab:wsj1}
\centering
\begin{tabular}{ll|l|l}
\hline
\multicolumn{2}{l|}{Number of Layers}               & \multicolumn{2}{l}{CER}    \\ \hline \hline
\multicolumn{1}{l|}{Self-attention} & Feed forward & eval92       & dev93        \\ \hline
\multicolumn{1}{l|}{12}             & 0            & 3.5          & 4.6          \\
\multicolumn{1}{l|}{11}             & 1            & \textbf{3.4} & \textbf{4.5} \\
\multicolumn{1}{l|}{10}             & 2            & 3.6          & 4.6          \\
\multicolumn{1}{l|}{9}              & 3            & 3.8          & 4.8          \\
\multicolumn{1}{l|}{8}              & 4            & 3.9          & 4.9          \\
\multicolumn{1}{l|}{7}              & 5            & 4.0          & 5.1          \\
\multicolumn{1}{l|}{6}              & 6            & 4.2          & 5.3          \\ \hline \hline
\multicolumn{1}{l|}{11}             & 0            & 3.6          & 4.7          \\
\multicolumn{1}{l|}{10}             & 0            & 4.0          & 5.2          \\ \hline \hline
\multicolumn{1}{l|}{12}             & 1            & 3.6          & 4.7          \\
\multicolumn{1}{l|}{11}             & 2            & 3.6          & 4.6          \\
\multicolumn{1}{l|}{11}             & 3            & 3.7          & 4.6          \\ \hline
\end{tabular}
\end{table}

To demonstrate that the range of learned context is stacked up and that multi-head attention is redundant for the upper layers of the encoder, we train models whose encoders are built by different numbers of self-attention layers and feed forward layers. For the encoder of these models, there are 12 layers in total and the lower layers are self-attention layers while the upper layers are feed forward layers. We start from an encoder with 6 self-attention layers and 6 feed forward layers. Then, we increase the number of self-attention layers and decrease the number of feed forward layers. Table~\ref{tab:wsj1} shows that as the number of self-attention layers increases, the character error rate (CER) decreases, which implies learning further contextual information is beneficial.

However, when the number of self-attention layers increases to 10, with 2 upper feed forward layers, the encoder gives almost identical results compared to the 12-layer self-attention baseline, although the 10-layer self-attention encoder has notably higher CERs. Furthermore, although the 11-layer self-attention encoder gives worse results compared to the 12-layer baseline, the encoder which has 11 self-attention layers and one upper feed forward layers yields the best results. Increasing the number of self-attention layers to 12 and decreasing the number of feed forward layers to 0 is harmful. This set of experiments shows the temporal relationships are well captured through 10 or 11 self-attention layers and further contextual information is unnecessary. Thus, feed forward layers are sufficient in learning further hidden representations. If the additional self-attention layer upon the 11 self-attention layers learns pure diagonal attention matrices, then it can be viewed as a feed forward layer. However, learning the diagonal attention is hard and redundant, whereas the feed forward layers guarantee ``diagonal attention". 

We tested if stacking more feedforward layers to make deeper encoders is beneficial. However, as shown in Table~\ref{tab:wsj1}, this does not give performance gains. We also tried to modify the architecture of the stacked feed forward layers, such as removing  residual connections or using an identity mapping \cite{he2016identity}. However, these modifications were not helpful and we did not observe CER reduction compared to the 11-layer self-attention 1-layer feed forward encoder.

\subsection{Experiments on SWBD}

We further test our argument on a larger and more complicated dataset, SWBD. The results are shown in Table~\ref{tab:SWBD}. The encoder with 10 self-attention layers have worse results than the encoders with 11 and 12 self-attention layers. Also, the 12-layer self-attention encoder has higher word error rates (WER) than the 11-layer encoder. However, the encoder with 10 self-attention layers and 2 feed forward layers, which has 12 layers in total, gives the lowest WERs. The 9 self-attention layers + 3 feed forward layers encoder yields worse WERs. Thus, the 10$^{th}$ self-attention layer is crucial in learning contextual information. Upon the 10$^{th}$ self-attention layer feed forward layers are sufficient in learning further abstract representations. 

\vspace{-2mm}
\begin{table}[h]
\caption{Word error rate (WER) of the experiments on SWBD. The evaluation sets are eval 2000 SWBD/callhome}
\vspace{-2mm}
\label{tab:SWBD}
\centering
\begin{tabular}{ll|l|l}
\hline 
\multicolumn{2}{l|}{Number of Layers}               & \multicolumn{2}{l}{WER}     \\ \hline \hline
\multicolumn{1}{l|}{Self-attention} & Feed forward & SWBD         & Callhome      \\ \hline
\multicolumn{1}{l|}{12}             & 0            & 9.0          & 18.1          \\
\multicolumn{1}{l|}{11}             & 1            & 9.0          & 17.8          \\
\multicolumn{1}{l|}{10}             & 2            & \textbf{8.9} & \textbf{17.6} \\
\multicolumn{1}{l|}{9}              & 3            & 9.5          & 18.5          \\ \hline \hline
\multicolumn{1}{l|}{11}             & 0            & 9.0          & 17.7          \\
\multicolumn{1}{l|}{10}             & 0            & 9.2          & 18.4          \\ \hline \hline
\multicolumn{2}{l|}{Transformer \cite{karita2019comparative}}                                                   & \multicolumn{1}{l|}{9.0}             &    18.1           \\ \hline
\end{tabular}
\end{table}

\vspace{-2mm}

\section{Conclusion}
In this paper, based on the argument that acoustic events often happen in small time spans with a strict left-to-right ordering and that the encoded context increases through the lowest self-attention layer to the highest self-attention layer through the Transformer encoder, we propose that for speech recognition, replacing the upper self-attention layers with feed forward layers will not harm the model's performance. Our experiments on WSJ and SWBD confirms this -- replacing the upper self-attention layers with feed forward networks even gives small error rate reductions. Future work include investigating this idea on other domains (e.g., natural language processing) and developing novel network architectures based on this observation.

\textbf{Acknowledgements:} Supported by a PhD Studentship funded by Toshiba Research Europe Limited (SZ),  and EPSRC Project SpeechWave, EP/R012180/1 (EL, PB, SR).

\bibliographystyle{IEEEtran}

\bibliography{mybib}


\end{document}